\documentstyle{article}
\begin{document}

\title{Are Strings Thermostrings?}

\author{Zahid Zakir \\
Institute of Noosphere, \\
167a B.Ipak Yuli, 700187 Tashkent, Uzbekistan}

\maketitle

\thispagestyle{empty}

\begin{abstract}
In the method of thermostring quantization the time evolution of point
particles at finite temperature kT is described in a geometric manner. The
temperature paths of particles are represented as closed (thermo)strings,
which are swept surfaces in space-time-temperature manifold. The method
makes it possible a new physical interpretation of superstrings IIA and
heterotic strings as point particles in a thermal bath with Planck
temperature.
\end{abstract}

\section{Introduction}

In the history of physics the replacement of local fundamental objects by
nonlocal ones has occurred numerous times. Newton's local particles with
nonlocal interactions were replaced by Maxwell's fields. In relativity
theory and in quantum physics point particles were reintroduced as
fundamental objects along with fields. Now we have strings as nonlocal
fundamental physical objects with local interactions \cite{GSW}. One direction
for the development of this theory would be to extend it into p-branes and
further abandon the locality of primary objects of the theory. Another
direction would be to restore the locality of primary objects by reducing
strings back to point particles and local fields with specific behavior.

The method of thermostring quantization \cite{ZZ1} \cite{ZZ2} is a simple step
in the latter direction. Here strings are considered as thermostrings, i.e.,
point particles in $(D-2)$-dimensional space with a thermal ''length''. Due
to the coupling of the coordinates of these particles with a temperature
degree of freedom, their time evolution in a D-dimensional
space-time-temperature manifold is described by the string formalism. This
method does not introduce any new objects into the physics. Instead it
considers new properties of old objects. The physical reason for nonlocality
is Gibbs ensemble averaging, and the length of the thermostring is nothing
but the thermal word line of the point particle \cite{Fe}.

A similar story happened in quantum theory with Schr$\ddot o$dinger's wave
mechanics. At first the wave function was considered to be a description of
real waves in physical space-time the same as electromagnetic waves. The
understanding that these are really ''probability waves'' of point particles
in configurational space was achieved through M. Born's statistical
interpretation of the wave function. Analogously, the thermostring
quantization is a statistical interpretation of strings. It explains the
appearance of the string behavior of particles at Planck distances naturally
by replacing the space-time foam with a thermal bath at Planck temperature.
In this approach space-time has a finite temperature, which can be taken
into account in a geometrical manner by introducing a temperature degree of
freedom as an additional dimension of the physical manifold.

\section{What is the String?}

In conventional string theory \cite{GSW}, strings are considered to be
fundamental one-dimensional objects in physical space. The introduction of
such nonlocal structures into physics leads to some conceptual difficulties
concerning the measurement of points of strings, intrinsic dynamics,
relativistic causality, etc. Strings are alien to the geometry of space-time
and the reduction of strings into space-time properties is a nontrivial
problem.

The M-theory attempts to construct strings in $10$ dimensions from geometry
of $11$ dimensional manifold, but this reduction does not change the
physical nature of strings. In $10$ dimensions we still have the old strings
as nonlocal objects in space.

\section{What is the Thermostring?}

The density matrix $\rho ({\bf r},\Delta \beta )$ for a nonrelativistic
particle at a finite temperature $kT=1/\Delta \beta ,$ after factorization $%
\Delta \beta =\beta -\beta _0$ can be represented as a transition amplitude $%
\rho ({\bf r},\beta ;{\bf r}_0,\beta _0)$ for ''pure'' states $\psi _i({\bf r%
},\beta )$ in the space-temperature manifold $({\bf r},\beta )$ \cite{Fe}:

\begin{equation}
\begin{array}{c}
\rho ( 
{\bf r},{\bf r}_0;\Delta \beta )=\sum \exp (-E_i\Delta \beta )\cdot \psi _i(%
{\bf r})\cdot \psi _i^{*}({\bf r}_0)= \\  \\ 
=\sum \psi _i({\bf r},\beta )\cdot \psi _i^{*}({\bf r}_0,\beta _0)=\rho (%
{\bf r},\beta ;{\bf r}_0,\beta _0) 
\end{array}
\end{equation}

where $\Delta \beta =\beta -\beta _0,$ $E_i$- is the energy of the particle, 
${\bf r}$ are $d$-dimensional space coordinates. Here wave functions $\psi
_i({\bf r},\beta )$ are the pure states of the particle with the Hamiltonian 
$H$ in $({\bf r},\beta )$-manifold:

\begin{equation}
\begin{array}{c}
\psi _i( 
{\bf r},\beta )=\exp (-H\beta )\cdot \psi _i({\bf r}),\psi _i^{*}(%
{\bf r},\beta )=\psi _i^{*}({\bf r})\cdot \exp (H\beta ) \\  \\ 
\int \psi _i^{*}({\bf r},\beta )\cdot \psi _j({\bf r},\beta )\cdot d{\bf r}%
=\delta _{ij} 
\end{array}
\end{equation}

and

\begin{equation}
\psi ({\bf r},\beta )=\int d{\bf r}_0\cdot \rho ({\bf r},\beta ;{\bf r}%
_0,\beta _0)\cdot \psi _i({\bf r}_0,\beta _0) 
\end{equation}

The partition function is defined as:

\begin{equation}
Z(\Delta \beta )=\int d{\bf r}\cdot \rho ({\bf r},\beta ;{\bf r}_0,\beta
_0)\mid _{{\bf r}(\beta )={\bf r}(\beta _0)} 
\end{equation}

The (inverse) temperature parameter of evolution $\beta $ here can be
considered as a geometrical dimension of the physical manifold in addition
to space and time. Here $\beta $ is limitless $-\infty \leq \beta \leq
\infty ,$ but the intervals $\Delta \beta $ are restricted by the condition $%
0\leq \Delta \beta \leq 1/kT.$

The density matrix as the $\beta $-evolution transition amplitude can be
represented through an infinity set of intermediate states \cite{Fe}:

\begin{equation}
\begin{array}{c}
\rho ( 
{\bf r},\beta ;{\bf r}_0,\beta _0)=\int d{\bf r}(\beta _1)...d{\bf r}(\beta
_n)\cdot \psi _i({\bf r},\beta )\cdot \psi _i^{*}({\bf r}_n,\beta _n)\psi _i(%
{\bf r}_n,\beta _n)\cdot ... \\  \\ 
...\cdot \psi _i^{*}({\bf r}_1,\beta _1)\psi _i({\bf r}_1,\beta _1)\cdot
\psi _i^{*}({\bf r}_0,\beta _0) 
\end{array}
\end{equation}

We can transfer all $\psi _i({\bf r}_k,\beta _k)$ to the left hand side and
all $\psi _i^{*}({\bf r}_k,\beta _k)$ to the right hand side. Then we can
compose wave functionals $\Psi _i,\Psi _i^{*}$ as the products of an
infinity number $(n\rightarrow \infty )$ intermediate state wave functions
of particles:

\begin{equation}
\begin{array}{c}
\rho ( 
{\bf r},\beta ;{\bf r}_0,\beta _0)=\int d{\bf r}(\beta _1)...d{\bf r}(\beta
_n)\cdot \{\psi _i({\bf r},\beta )\cdot \psi _i({\bf r}_n,\beta _n)...\psi
_i({\bf r}_1,\beta _1)\}\cdot \\  \\ 
\cdot \{\psi _i^{*}( 
{\bf r}_n,\beta _n)...\psi _i^{*}({\bf r}_1,\beta _1)\cdot \psi _i^{*}({\bf r%
}_0,\beta _0)\}={\sum }{\int\limits_{{\bf r}(\beta _0)}^{{\bf r}(\beta )}}D%
{\bf r}(\beta ^{\prime })\cdot \Psi _i[{\bf r}(\beta ^{\prime });\beta
,\beta _0]\cdot \Psi _i^{*}[{\bf r}(\beta ^{\prime });\beta ,\beta _0] \\  
\end{array}
\end{equation}

Here $D{\bf r}(\beta )$ is the path integration measure. The wave functional 
$\Psi $ describes the states the temperature path:

\begin{equation}
\begin{array}{c}
\Psi _i[ 
{\bf r}(\beta ^{\prime });\beta _n,\beta _1]=\psi _i({\bf r}_n,\beta _n)\psi
_i({\bf r}_{n-1},\beta _{n-1})...\psi _i({\bf r}_1,\beta _1) \\  \\ 
\int D{\bf r}(\beta ^{\prime })\cdot \Psi _i^{*}[{\bf r}(\beta ^{\prime
});\beta ,\beta _0]\cdot \Psi _j[{\bf r}(\beta ^{\prime });\beta ,\beta
_0]=\delta _{ij} 
\end{array}
\end{equation}

We see that the temperature path of the point particle ($\beta -$world line)
can be represented as some one dimensional physical object in
space-time-temperature manifold with the wave functional $\Psi $. We will
call this object as a {\it thermostring} \cite{ZZ1}.

In the general case, the wave functional $\Psi $ must be symmetrized under
permutations of points of the temperature paths or thermostrings. These
permutations depend on the type of statistics of the particles in the Gibbs
ensemble (bosonic or fermionic) and they determine the type of statistics of
thermostrings. In ordinary string theory such permutations are impossible
since ordinary strings are treated as continuous one dimensional objects in
physical space.

The time evolution of the density matrix can be described by summing all of
the surfaces which are swept by the temperature path in time. This
circumstance allow us to introduce the thermostring representation of the
quantum statistical mechanics of particles or {\it the thermostring
quantization} of particles at finite temperatures.

Taking into account time evolution formulas for wave functions $\psi _i({\bf %
r},\beta ):$

\begin{equation}
\psi _i({\bf r},\beta ,t)=\exp (-iHt)\cdot \psi _i({\bf r},\beta ), 
\end{equation}

we can obtain the time evolution expression for the wave functional $\Psi .$

In the $({\bf r},\beta )-$ manifold we have $(d+1)$-dimensional coordinates $%
{\bf q}$ with components $({\bf r},\beta ).$ Time derivatives of these
vectors are space-temperature velocities of particles. They are separated
into longitudinal components and transverse to the temperature path
components:

\begin{equation}
\begin{array}{c}
\partial 
{\bf q}/\partial t={\bf v=v}_{\perp }+{\bf v}_{\parallel } \\  \\ 
{\bf v}_{\perp }={\bf v}-{\bf k}\cdot ({\bf q}^{\prime }\cdot {\bf v}),{\bf q%
}^{\prime }=\partial {\bf q}/\partial \beta ,{\bf k}={\bf q}^{\prime }/{\bf q%
}^{\prime 2} 
\end{array}
\end{equation}

The longitudinal components of ${\bf v}_{\parallel }$ also have two parts.
The first part leads to the collective motion of the thermostring as a whole
object with synchronous displacements of all points of the thermostring. In
the case of a single thermostring this displacements can be disregarded as
the zero mode. The second part of the longitudinal velocity leads to
permutations of the points of the thermostring, eg., the replacement of
neighboring points. This permutation does not contribute to the energy of
the thermostring because of the indistinguishability of the particles in
Gibbs ensemle. In ordinary string theory the exclusion of ${\bf v}%
_{\parallel }$ from the Lagrangian is one of the difficulties of the theory 
\cite{BN}, whereas in the case of thermostrings this is a natural and
necessary process.

So, we have following time evolution expression for the wave functional:

\begin{equation}
\Psi _i[{\bf q}(\overline{\beta },t),\beta ,\beta _0;t]=\exp \{-\frac{%
i(t-t_0)}{\Delta \beta }{\int\limits_{\beta _0}^\beta }d\beta ^{\prime
}\cdot H({\bf v}_{\perp }^2)\}\cdot \Psi _i[{\bf q}(\beta ^{\prime });\beta
,\beta _0;t_0] 
\end{equation}

The action function for thermostrings is:

\begin{equation}
S[{\bf q}]=\frac 1{\Delta \beta }\int d\beta dtL[{\bf v}_{\perp }(\beta ,t),%
{\bf q}(\beta ,t)]=\frac m{2\Delta \beta }\int d\beta dt\cdot {\bf v}_{\perp
}^2 
\end{equation}

In the case of the Gibbs ensemble of free relativistic particles we have the
following action function for corresponding relativistic thermostrings:

\begin{equation}
S[x]=-\frac m{\Delta \beta }\int d\beta dt\cdot \sqrt{1-{\bf v}_{\perp }^2} 
\end{equation}

where $(d+2)$-vector $x^\mu $ have the components $x^\mu ({\bf q,}t)=x^\mu (%
{\bf r,}\beta ,t{\bf ).}$ It can be shown \cite{BN} that after introduction
of world sheet parameters $\tau ,\sigma $ and substitutions:

\begin{equation}
\begin{array}{c}
t=t(\tau ,\sigma ), 
d\beta =d\sigma \sqrt{x^{\prime 2}},\stackrel{.}{x}^\mu
=\partial x^\mu /\partial \tau ,x^{\prime }=\partial x/\partial
\sigma , \\  \\ 
dtd\sigma = 
\frac{\partial (t,\sigma )}{\partial (\tau ,\sigma )}d\tau d\sigma =%
\stackrel{.}{t}\cdot d\tau d\sigma , \\  \\ 
\partial {\bf q/}\partial t=\stackrel{.}{\bf q}/\stackrel{.}{t},%
\partial {\bf q/}\partial \sigma ={\bf q}^{\prime }-\stackrel{.}{\bf q}\cdot
(t^{\prime }/\stackrel{.}{t}), 
\end{array}
\end{equation}

this expression leads to the Nambu-Goto action for the relativistic
thermostring:

\begin{equation}
S[x]=-\gamma {\int }d\sigma d\tau \cdot \sqrt{(\stackrel{.}{x{\bf \cdot }}%
x^{\prime })^2-\stackrel{.}{x}^2\cdot x^{\prime 2}} 
\end{equation}

Here $\gamma =m/\Delta \beta ,$ $\sigma $ and $\tau $ are the world sheet
coordinates of thermostring. We see that in the thermostring representation
of the quantum statistical mechanics of particles there exist
reparametrizational symmetry $\sigma ^{\prime }=f(\sigma ,\tau ),$ $\tau
^{\prime }=\varphi (\sigma ,\tau ),$ just as in string theories. This action
is fully relativistically invariant if $\Delta \beta $ and the limits of $%
\beta $-integration are invariants. At ordinary temperatures this is
impossible, but if we take the relativistically invariant Planck temperature 
$T_p$ as the limiting temperature of thermostrings with $\Delta \beta
=1/kT_p $, we have an invariant action function.

The action function and the periodicity conditions $x^\mu (\sigma ,\tau
)=x^\mu (\sigma +\pi ,\tau )$ are leads to the equations for $x_\mu (\sigma
,\tau )$ with solutions:

\begin{equation}
\begin{array}{c}
x^\mu (\sigma ,\tau )=x_R^\mu (\sigma ,\tau )+x_L^\mu (\sigma ,\tau ) \\  
\\ 
x_{R,L}^\mu (\sigma ,\tau )=\frac 12x_0^\mu +\frac 12l^2p^\mu (\tau \mp
\sigma )+\frac{il}2{\sum }\frac 1n\alpha _{n,\pm }^\mu e^{-2in(\tau \mp
\sigma )} 
\end{array}
\end{equation}

where $n\neq 0$, $x_0^\mu $, $p^\mu $- constants, $l^2=1/\pi \gamma $. Then
the operators for dynamical variables can be constructed and they are leads
to the Virasoro algebra and to the spectrum of states identical with the
closed bosonic string case. In the case of the fermionic particles the
fermionic thermostrings with the periodicity conditions for fermionic
degrees of freedom can be obtained.

We can describe the time evolution by surface integration along the world
surface which the thermostrings are swept in time. The result is identical
with the Polyakov surface integral for closed strings.

\section{Are Strings Thermostrings?}

We can replace the space-time foam at Planck distances with a thermal bath
with Planck temperature. When we take the influence of this thermal bath on
the particle's behavior into account, we come to the thermostring
quantization scheme. At Planck distances particles must be described as
thermostrings and we can interpret superstrings as thermostrings, i.e., as
point particles in a Planckian thermal bath. This interpretation preserves
all of the achievements of ordinary string theory and at the same time
excludes some of the conceptual difficulties associated with the
introduction of nonlocal objects with unmeasurable intrinsic structure.

Thermostring interpretation of strings leads to the following general
consequences:

a) One of the dimensions of the string theory manifold is the (inverse)
temperature dimension and therefore the dimension of space is $d=8$ which,
together with the time and temperature degrees of freedom, combine into the
critical dimension $D=d+2=10$. This fact is important in the
compactification of six dimensions which can be represented as $5+1$ (5
space and 1 temperature).

b) At the initial and final states in the amplitudes of strings, only closed
strings can appear as observable physical states.

c) The charge of particles must be distributed along the length of the
thermostring.

Among superstring theories only theories of closed strings satisfy these
conditions and we may conclude that only superstrings IIA (in the case of
neutral particles) and heterotic strings (in the case of charged particles) 
\cite{GSW} can be interpreted as thermostrings. This means that the
thermostring interpretation selects only one theory for each type of
particle from the family of string theories.

We can describe thermostring interactions simultaneously in particle,
statistical ensemble, and string languages. The factorization of one
statistical ensemble into two ensembles or the merging of two ensembles into
one are physically clearer and simple procedures than the cutting or gluing
of rigid strings in conventional string theory. We can perform the
thermostring quantization of physical strings as one-dimensional objects in
physical space and as a result obtain a theory of membranes. If we perform
the thermostring quantization of physical p-branes we shall obtain a theory
of (p+1)-branes, i.e., the dimensionality of initial objects of the theory
increases by one. In the treatment of p-branes the thermostring
representation can be combined with M-Theory methods if we interpret one of
its eleven dimensions as temperature.

Thus, the thermostring quantization can be the simplest and natural way to
understand the modification of local theories at Planck distances and
temperatures.

\end{document}